\documentclass[11pt,fleqn]{article}
\usepackage{latexsym}
\usepackage{amssymb}
\usepackage{amsmath}
\usepackage[hypertex]{hyperref}
\usepackage{graphicx}

\usepackage{slashed}

\makeatletter 
\newcommand\fleqnoff{\@fleqnfalse\@mathmargin\@centering} 
\newcommand\fleqnon[1][\leftmargini]{\@fleqntrue\@mathmargin=#1\relax 
  \@ifundefined{mathindent}{\let\mathindent\@mathmargin}{}} 
\makeatother 

\begin{document}
\pagestyle{empty}

\begin{flushright}
\today \\
TU-893
\end{flushright}

\vspace{3cm}

\begin{center}

{\bf\LARGE Spectral-Function Sum Rules \vspace{3mm}\\
in Supersymmetry Breaking Models}
\\

\vspace*{1.5cm}
{\large 
Ryuichiro Kitano$^a$, Masafumi Kurachi$^b$, Mitsutoshi Nakamura$^a$, and Naoto Yokoi$^a$
} \\
\vspace*{0.5cm}

$^a${\it Department of Physics, Tohoku University, Sendai 980-8578, Japan}\\
$^b${\it Kobayashi-Maskawa Institute for the Origin of Particles and the Universe\\ Nagoya University, Nagoya 464-8602, Japan}\\
\vspace*{0.5cm}

\end{center}

\vspace*{1.0cm}

\begin{abstract}
{\normalsize

The technique of Weinberg's spectral-function sum rule is a powerful tool 
for a study of models in which global symmetry is
dynamically broken. It enables 
us to convert information on the short-distance behavior of a theory to relations 
among physical quantities which appear in the low-energy picture of the theory.
We apply such technique to general supersymmetry breaking models to derive 
new sum rules.

}
\end{abstract}

\newpage
\baselineskip=18pt
\setcounter{page}{2}
\pagestyle{plain}
\baselineskip=18pt
\pagestyle{plain}
\fleqnoff

\setcounter{footnote}{0}

\section{Introduction}

In theories with spontaneously broken global symmetry, the infrared (IR)
physics is described by the Nambu-Goldstone particle(s) and their
interactions are restricted by the broken (and unbroken)
symmetries. Those restrictions are generically called the low-energy
theorems and apply to any models of symmetry breaking.

There is a less model-independent but powerful non-perturbative result
called the Weinberg sum rules~\cite{Weinberg:1967kj}.
These are relations among spectral
functions, and can be derived if the ultraviolet (UV) theory is
asymptotically free and the symmetry is broken by a vacuum expectation
value (VEV) of an operator whose mass dimension is high enough.
The ingredients for deriving the Weinberg sum rules are (well-defined)
operators such as currents and their transformation laws under the
broken symmetry.
In the case of chiral symmetry breaking in QCD, by using the
charge-current algebra, Weinberg has derived two sum rules.
Once the spectral functions are approximated
by summation of one-particle states of hadrons, the rules reduce to
relations among hadron masses and decay constants: $ f_\pi^2 - f_\rho^2
+ f_{a_1}^2 \simeq 0$ and $ m_\rho^2 f_\rho^2 - m_{a_1}^2 f_{a_1}^2
\simeq 0$. They catch qualitative features of hadron properties
correctly.

In this paper, we apply such procedure to the case of dynamical 
supersymmetry (SUSY) breaking, and derive new sum rules among 
physical quantities in several models.
Those sum rules are predictions of the dynamical SUSY breaking models,
and even apply to the ``incalculable models,'' such as the models
proposed in Ref.~\cite{Affleck:1983vc,Affleck:1984mf}.
If there is a weakly coupled 
description of hadrons at low energy, {\it
i.e.}, the dual theory, the sum rules reduce to approximate relations
among masses and decay constants. These relations can be used as a
window between the UV and IR descriptions of dynamical SUSY breaking
models.

Our approach is related to the study in Ref.~\cite{Meade:2008wd}, where
a technique is developed to describe soft SUSY breaking parameters in terms
of current correlators in the hidden sector.
The formulation, called the general gauge mediation, was used in various
contexts, such as in models with gauge
messengers~\cite{Intriligator:2010be}.
Recently, Ref.~\cite{Fortin:2011ad} discussed a way to calculate the
current correlators by using the operator product expansion~(OPE)
in approximately superconformal theory.

In the next section, we apply the Weinberg's method to the direct gauge
mediation models. With sum rules derived there, we show that the
sfermion mass squared is expressed in terms of masses of the spin 0, 1/2
and 1 particles in the SUSY breaking sector.  Then, in
section~\ref{sec:supercurrent}, we use the same technique to extract the
sum rules which can be derived from several correlators of components in
the supercurrent multiplet.  The supercurrent multiplet is known to be
well-defined in a wide class of SUSY theories. From the transformation
laws of the component fields, a set of sum rules can be derived
involving states with spins $0$, $1/2$, $1$, $3/2$, and $2$.

\section{Direct gauge mediation and sum rules}
\label{sec:GGM}

In this section, by using the language of the general gauge
mediation~\cite{Meade:2008wd}, we derive sum rules which are related to 
the current correlators in the hidden sector. Then, with those sum
rules, we show that the sfermion mass squared can be expressed in terms
of masses of the spin 0, 1/2 and 1 particles in the SUSY breaking
sector.

\subsection{Current multiplet and correlators}

We introduce the current superfield ${\cal J}={\cal J}(x, \theta, \bar{\theta})$. 
It is defined as a real linear superfield which satisfies the current conservation 
conditions, $\bar{D}^2 {\cal J}=D^2 {\cal J}=0$. In components, it can be expressed 
as
\begin{eqnarray}
 {\cal J} = J + i \theta j - i \bar \theta \bar j
+ \theta \sigma^\mu \bar \theta j_\mu
- {1 \over 2} \theta \theta \bar \theta \bar \sigma^\mu \partial_\mu j
+ {1 \over 2} \bar \theta \bar \theta 
\theta \sigma^\mu \partial_\mu \bar j
- {1 \over 4} \theta \theta \bar \theta \bar \theta \Box J.
\end{eqnarray}
Transformation laws of these component fields under SUSY are given by
\begin{eqnarray}
 \delta_Q J &=& -i \eta j,\\
 \delta_Q j_\alpha &=& 0,\\
 \delta_Q \bar j^{\dot \alpha} &=& i (\bar \sigma^\mu \eta)^{\dot \alpha} (j_\mu + i \partial_\mu J),\\
 \delta_Q j_\mu &=& - \eta \partial_\mu j.
\end{eqnarray}
Here, $\eta$ is a parameter of the SUSY transformation, and 
we defined $\delta_Q
{\cal O} = - \eta_\alpha \delta_Q^\alpha {\cal O}$. 

Now, we consider the 
following current correlators\footnote{We define $\langle \cdots \rangle$ by the
path integral, and thus they are Lorentz covariant.}:
\begin{eqnarray}
 D_1^{\mu \nu} (x,y)
&\equiv&
\Big \langle 
\delta_Q^\alpha \left[
\bar j^{\dot \alpha} (x)
j^{\mu} (y)
\right]
\Big \rangle  
(\sigma^\nu)_{\alpha \dot \alpha},\label{eq:D1}\\
D_2^{\mu} (x,y)
&\equiv&
\Big \langle 
\delta_Q^\alpha \left[
\bar j^{\dot \alpha} (x)
J (y)
\right]
\Big \rangle  
(\sigma^\mu)_{\alpha \dot \alpha}.\label{eq:D2}
\end{eqnarray}
These $D$'s should vanish if SUSY is unbroken.
For later convenience, we rewrite Eqs.~(\ref{eq:D1}) and (\ref{eq:D2})
in terms of the Fourier transformed functions, $\tilde{C}$'s, introduced
in Ref.~\cite{Meade:2008wd}:
\begin{eqnarray}
	\tilde C_0(k^2) &=& 
		\int\frac{d^4x}{i(2\pi)^4}\Big\langle J(x)J(y)\Big\rangle e^{ik\cdot(x-y)}, \\
	-(\bar{\sigma}_\mu)^{\dot\alpha\alpha}k^\mu\tilde C_{1/2}(k^2) &=& 
		\int\frac{d^4x}{i(2\pi)^4}\Big\langle j^\alpha(x)\bar{j}^{\dot\alpha}(y)\Big\rangle e^{ik\cdot(x-y)},\\
	-(k^2\eta^{\mu\nu} -k^\mu k^\nu )\tilde C_1(k^2) &=& 
		\int \frac{d^4x}{i(2\pi)^4}\Big\langle j^\mu(x) j^\nu(y)\Big\rangle e^{ik\cdot (x-y)},\\
	\epsilon_{\alpha\beta} M \tilde B_{1/2}(k^2) &=& 
		\int \frac{d^4x}{i(2\pi)^4}\Big\langle j_\alpha(x) j_\beta(y)\Big\rangle e^{ik\cdot (x-y)},
\end{eqnarray}
where $M$ is a characteristic mass scale of the theory.
Using those $\tilde{C}$'s, we can write down the $k^\mu k^\nu$ part 
($k^\mu$ part) of $D_1$ ($D_2$) as follows:
\begin{eqnarray}
 D_1^{\mu \nu} |_{k^\mu k^\nu}
&=& - 2 i \int {d^4 k \over i (2 \pi)^4} k^\mu k^\nu
\left(
\tilde C_1 (k^2) - \tilde C_{1/2} (k^2)
\right) e^{-i k \cdot (x-y)},\\
 D_2^{\mu} |_{k^\mu}
&=& - 2 i \int {d^4 k \over i (2 \pi)^4} k^\mu
\left(
\tilde C_0 (k^2) - \tilde C_{1/2} (k^2)
\right) e^{-i k \cdot (x-y)}.
\end{eqnarray}

\subsection{Weinberg sum rules}
\label{sec:WS}
\begin{figure}[t]
\begin{center}
 \includegraphics[width=7cm]{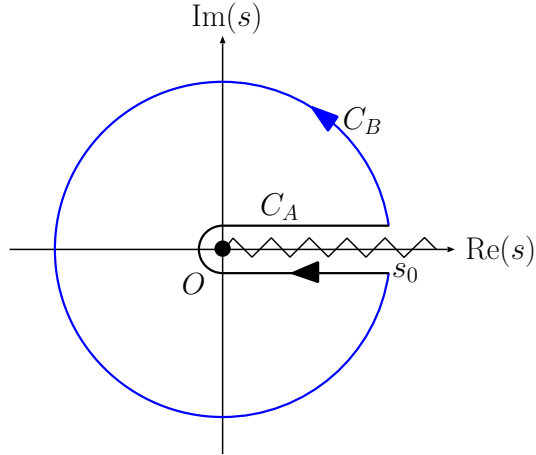}
\end{center}
\caption{Contour of the integral.}
\label{fig:contour}
\end{figure}

In this subsection, we explain the procedure of deriving sum rules rather 
in detail.
We use $D_1$ as an example in the following discussion. Let us define
\begin{eqnarray}
 \Pi_{D_1} (s) \equiv \tilde C_1 (s) - \tilde C_{1/2} (s),
\end{eqnarray}
and extend the function $\Pi_{D_1}$ to a complex plane; it has a
branch cut on the real and positive value of $s$.
By the Cauchy integral theorem, we obtain the following identity:
\begin{eqnarray}
0 = 
\int_{C_A} ds \  s^n \Pi_{D_1} (s)
+ \int_{C_B} ds \  s^n \Pi_{D_1} (s),
\label{eq:id}
\end{eqnarray}
where $n$ is an integer. The paths $C_A$ and $C_B$ are shown in
Fig.~\ref{fig:contour} where $s_0$ is an arbitrary real and positive
number.

The Weinberg sum rules can be obtained by using the OPE
for the second integral. 
We here consider an asymptotically free theory where the OPE at a UV
scale can be done perturbatively.
Let ${\cal O}$ be the lowest dimensional operator whose VEV breaks SUSY,
and $d$ be the mass dimension (defined by the classical scaling in the
UV theory) of ${\cal O}$.
Since $\Pi_{D_1}$ is dimensionless, it can be expanded as
\begin{eqnarray}
 \Pi_{D_1} (s) 
\simeq {c_{\cal O} \langle {\cal O} \rangle \over (-s)^{d/2}} + \cdots,
\label{eq:expansionOfCorrelator}
\end{eqnarray}
where $\cdots$ are higher order terms in the $1/(-s)$ expansion and
$c_{\cal O}$ is a dimensionless coefficient. 
Here $d/2$ should be an integer since it can be obtained by a
calculation of Feynman diagrams. If $d/2$ is not an integer, such an
operator either does not contribute or should be supplied by some
dimensionful parameter in the Lagrangian.
The second integral in Eq.~(\ref{eq:id}) vanishes for $n <
d/2-1$.
In general, $d \le 4$ since $T^{\mu}_\mu$ can always be ${\cal
O}$.

On the other hand, the function $\Pi_{D_1} (s)$ for the real and
positive $s$ can be expressed in terms of a spectral function as
follows:
\begin{eqnarray}
 \Pi_{D_1} (s) = 
- \int_0^{\infty} d\sigma^2 
{
\rho_{D_1} (\sigma^2) 
\over s -
  \sigma^2  + i \epsilon}
+ \Delta (s),
\label{eq:spectral}
\end{eqnarray}
where $\Delta (s)$ represents contact terms which are regular
everywhere.
By using the expression in Eq.~(\ref{eq:spectral}), the first integral
in Eq.~(\ref{eq:id}) reduces to
\begin{eqnarray}
2 \pi i
 \int_0^{s_0} ds \  s^n \rho_{D_1}(s)
\label{eq:firstint}
\end{eqnarray}
for $n \ge 0$. For $n<0$, the integral depends on $\Delta (s)$.

In asymptotically free theories, the use of the OPE is justified when
$(-s)$ is sufficiently large. Therefore, the quantity (\ref{eq:firstint})
should asymptotes to zero for $s_0 \to \infty$ if $n$ is within the
window:
\begin{eqnarray}
0 \le n < {d \over 2} -1.
\label{eq:n-region}
\end{eqnarray}
  From $d \le 4$, such $n$ can only be zero. In summary, we obtain
\begin{eqnarray}
 \int_0^{\infty} ds \ \rho_{D_1} (s) = 0,
\label{eq:sum-spectral}
\end{eqnarray}
for models with $d=3$ or $4$.

The sum rules we obtain from $D_1$ and $D_3$ are
\begin{eqnarray}
 \int_0^{\infty} ds \ \left(\rho_1 (s) - \rho_{1/2} (s) \right)&=& 0,\label{GGM:sum1}\\
 \int_0^{\infty} ds \ \left(\rho_0 (s) - \rho_{1/2} (s) \right)&=& 0,\label{GGM:sum2}
\end{eqnarray}
where
\begin{eqnarray}
 \tilde C_a (s) = - \int_0^\infty d\sigma^2 {\rho_a (\sigma^2) \over s -
  \sigma^2 + i\epsilon}.\ \ \ \ (a=0,\ 1/2,\ 1)
\end{eqnarray}
No sum rule for $\tilde B_{1/2}$ is obtained from other
correlators.

\subsection{Low energy models and sum rules}

Let us assume that the SUSY breaking model is a confining theory and 
its low-energy physics is well described by the lowest modes {\it \`a la}
Weinberg~\cite{Weinberg:1967kj}:
\begin{eqnarray}
 \rho_a (s) = f_a^2 \delta (s - m_a^2).
 \label{eq:rho}
\end{eqnarray}
In this case, the sum rules Eqs.~\eqref{GGM:sum1} and \eqref{GGM:sum2} suggest
\begin{eqnarray}
 f_0^2 = f_{1/2}^2 = f_1^2 \equiv f^2_h.
\label{eq:f}
\end{eqnarray}
It states that the decay constants are the same even though the masses can split.

By using the formula of the general gauge mediation~\cite{Meade:2008wd},
the scalar masses via gauge mediation are given by
\begin{eqnarray}
 m_s^2 &=&  g^4 c_2 
\int {d^4 k \over i (2 \pi)^4} {1 \over k^2}
\left(
3 \tilde C_1 (k^2)
- 4 \tilde C_{1/2} (k^2)
+ \tilde C_0 (k^2)
\right)
\nonumber \\
&=& {g^4 c_2 f^2_h \over (4\pi)^2} \log {m_0^2 m_1^6 \over m_{1/2}^8}.
\end{eqnarray}
Here, $m_0$, $m_{1/2}$, $m_{1}$ are masses of the particles with spin 0,
1/2, and 1 in the hidden sector, respectively, and $c_2$ is the
quadratic Casimir invariant. A finite result is obtained due to the sum
rules. (Similar to the $\pi^+ - \pi^0$ mass splitting by
QED. See~\cite{Das:1967it}.)  Interestingly, in
Ref.~\cite{Intriligator:2010be}, the same expression for the sfermion
mass squared was derived in a model with gauge messengers.

From Eqs. (\ref{eq:rho}) and (\ref{eq:f}), 
the gaugino masses can also be calculated as  
\begin{eqnarray}
 m_{\lambda} = {g^2 f^2_h \over m_{1/2}}.
\end{eqnarray}
In summary, by using the sum rules, one can express the sfermion and
gaugino masses by hadron masses in the hidden sector.

\section{Supercurrent and sum rules}
\label{sec:supercurrent}

In this section, as another example, we apply the same procedure used in the previous 
section to the supercurrent multiplet of the SUSY breaking sector.

\subsection{Supercurrent and correlators}

In a wide class of supersymmetric field theories, one can define a real
supermultiplet called the supercurrent ($J^\mu$)~\cite{Ferrara:1974pz}
(See~\cite{Komargodski:2010rb} for a recent discussion).  It is composed
of the SUSY current ($S^\mu_\alpha$), the symmetric energy momentum
tensor ($T^{\mu \nu}$), the $R$-current ($j^\mu$), and a scalar operator
$x$.
The $R$-current defined in this way is not conserved unless the theory
is conformal. 
The transformation laws of those component fields under SUSY are given
by
\begin{eqnarray}
 \delta_Q j_\mu &=&
- i \eta \left(
S_\mu - {1 \over 3} \sigma_\mu \bar \sigma^\nu S_\nu
\right),\label{SUSY:Tr1}\\
 \delta_Q x &=& -{ 2 \over 3 }  i \eta \sigma^\mu
  \bar S_\mu,\\
 \delta_Q x^\dagger &=& 0,\\
 \delta_Q S_{\mu \alpha}
&=& 2 (\sigma_{\mu \nu} \eta )_\alpha \partial^\nu x^\dagger,\\
 \delta_Q \bar S_{\mu}^{\dot \alpha}
&=& i (\bar \sigma^\nu \eta)^{\dot \alpha}
\left[
2 T_{\mu \nu} + i \partial_\nu j_\mu 
- i \eta_{\mu \nu} \partial \cdot j
- {1 \over 2} \epsilon_{\mu \nu \rho \sigma} \partial^\rho j^\sigma
\right],\\
 \delta_Q T^{\mu \nu} &=&
- {1 \over 2}\left[
\eta \sigma^{\rho \mu} \partial_\rho S^\nu
+ 
\eta \sigma^{\rho \nu} \partial_\rho S^\mu
\right]\label{SUSY:Tr6}.
\end{eqnarray}

By using the above component fields, 
we define the following set of current correlators:
\begin{eqnarray}
C_1^{\mu \nu \rho \sigma} (x,y)
&\equiv&
\Bigg \langle 
\delta_Q^\alpha \left[
\bar S^{\mu \dot \alpha} (x)
T^{\rho \sigma} (y)
\right]
\Bigg \rangle  
(\sigma^\nu)_{\alpha \dot \alpha},\label{currentCorrelatorC1}\\
C_2^{\mu \nu \rho \sigma \kappa} (x,y)
&\equiv&
\Bigg \langle 
\delta_Q^\alpha \left[
S^{\mu}_\gamma (x)
T^{\rho \sigma}
\right]
\Bigg \rangle  
(\sigma^{\nu \kappa})_{\alpha}^{\ \gamma},\\
C_3^{\mu \nu}  (x,y) &\equiv&
\Bigg \langle 
\delta_Q^\alpha \left[
\bar S^{\mu \dot \beta} (x)
x^\dagger (y)
\right]
\Bigg \rangle
(\sigma^\nu)_{\alpha \dot \beta},\\
C_4^{\mu \nu \kappa} (x,y) &\equiv&
\Bigg \langle 
\delta_Q^\alpha \left[
S^{\mu}_\gamma (x)
x^\dagger (y)
\right]
\Bigg \rangle  
(\sigma^{\nu \kappa})_{\alpha}^{\ \gamma},\\
C_5^{\mu \nu} (x,y) &\equiv&
\Bigg \langle 
\delta_Q^\alpha \left[
\bar S^{\mu \dot \beta} (x)
x (y)
\right]
\Bigg \rangle  
(\sigma^{\nu})_{\alpha \dot \beta},\\
C_6^{\mu \nu \kappa} (x,y) &\equiv&
\Bigg \langle 
\delta_Q^\alpha \left[
S^{\mu}_\gamma (x)
x (y)
\right]
\Bigg \rangle  
(\sigma^{\nu \kappa})_{\alpha}^{\ \gamma},\\
C_7^{\mu \nu \rho} (x,y) &\equiv&
\Bigg \langle 
\delta_Q^\alpha \left[
\bar S^{\mu \dot \beta} (x)
j^\rho (y)
\right]
\Bigg \rangle  
(\sigma^{\nu})_{\alpha \dot \beta},\\
C_8^{\mu \nu \rho \kappa} (x,y) &\equiv&
\Bigg \langle 
\delta_Q^\alpha \left[
S^{\mu}_\gamma (x)
j^\rho (y)
\right]
\Bigg \rangle  
(\sigma^{\nu \kappa})_{\alpha}^{\ \gamma}\label{currentCorrelatorC8}.
\end{eqnarray}
If SUSY is unbroken, all of them are vanishing. 

Since it will become important when we derive sum rules, let us here
discuss $R$-charges associated with the above correlators.  The
$R$-symmetry plays a crucial role for SUSY
breaking~\cite{Nelson:1993nf}, and in most cases, it is assumed that UV
theories of SUSY breaking models are $R$-symmetric.  Therefore, in the
present study, we assume that UV theories, from which OPE of the
correlators are calculated, have $R$-symmetry.  The $R$-charges
associated with each correlator are uniquely fixed since the components
of the supercurrent have $R$-charges determined from the SUSY algebra.
Those are summarized in Table~\ref{tab:R-charge},
\begin{table}[t]
\begin{center}
\begin{tabular}{|c|c|c|}
\hline
Correlators & $R$-charge & Dim. of ${\cal O}$ \\ \hline \hline
$C_{1},~ C_{6},~ C_{7}$ & $0$ & $d_{0}$ \\ \hline
$C_{2},~ C_{3},~ C_{5},~ C_{8}$ & $2$ &  $d_{2}$ \\ \hline
$C_{4}$ & $4$ & $d_{4}$ \\ \hline
\end{tabular}
\end{center}
\caption{$R$-charges associated with each correlator.  
$d_0$, $d_2$ and $d_4$ denote the dimension of the lowest-dimension 
SUSY breaking operator which contribute to the OPE of correlators with 
$R=0$, $2$ and $4$.}
\label{tab:R-charge}
\end{table}
and operators that appear in the OPE of each correlator should have the 
same $R$-charges as corresponding correlators. 
If the $R$-symmetry is 
not broken spontaneously, correlators with non-zero $R$-charges should 
vanish identically, and only correlators with zero $R$-charge, namely 
$C_1$, $C_6$ and $C_7$, would provide non-trivial sum rules. 
Meanwhile, if the $R$-symmetry is spontaneously broken, 
correlators with non-zero $R$-charges are also non-vanishing, and 
further sum rules can be derived. For later convenience, we introduce 
$d_0$, $d_2$ and $d_4$ to denote the dimension of the lowest-dimension 
SUSY breaking operator which contribute to the OPE of correlators with 
$R=0$, $2$ and $4$. (See Table~\ref{tab:R-charge}.) 
The number of sum rules we can 
derive from each correlator depends on values of $d_{0,2,4}$ as we 
will discuss in detail later.

\subsection{Sum rules in effective theories}

An explicit form of sum rules can be derived by approximating the
spectral function by one-particle states of hadrons.
Such an approximation is valid when there is a weakly coupled
description of hadrons at low energy.
We assume that there is such an effective description.
As hadronic degrees of freedom, we introduce fields with spins from 0 to
2 as follows:
\begin{itemize}
 \item $\phi$ (massive or massless spin 0 (scalar)),
 \item $\pi$ (massive or massless spin 0 (pseudoscalar)),
 \item $\lambda$ (the Goldstino, spin 1/2, massless),
 \item $\chi$ (massive spin 1/2 (Majorana)),
 \item $v_\mu$ (massive spin 1 (real)),
 \item $\psi_\mu$ (massive spin 3/2),
 \item $h^{\mu \nu}$ (massive spin 2).
\end{itemize}
Except for $\lambda$, there can exist multiple particles with the same spin
and parity.
In the following, we suppress the indices associated with such multiple
particles. The sum rules we obtain below should be
understood as the one with summations of these indices.

One particle parts of the supercurrent multiplet can be parametrized as
follows:
\begin{eqnarray}
 S^\mu_{\alpha} = 
i f^4 \sigma^\mu \bar \lambda 
- 2 f^2 f^\prime \sigma^{\mu \nu}  \partial_\nu \lambda
- 2 m_\psi f_{\psi} \sigma^{\mu \nu} \psi_\nu
- 2 f_{\chi} \sigma^{\mu \nu} \partial_\nu \chi + \cdots,
\end{eqnarray}
\begin{eqnarray}
 \bar S^{\mu \dot \alpha} 
= i f^4 \bar \sigma^\mu \lambda 
- 2 f^2 f^{\prime *} \bar \sigma^{\mu \nu}  \partial_\nu \bar \lambda
- 2 m_\psi f_{\psi}^* \bar \sigma^{\mu \nu} \bar \psi_\nu
- 2 f_{\chi}^* \bar \sigma^{\mu \nu} \partial_\nu \bar \chi + \cdots,
\end{eqnarray}
\begin{eqnarray}
 T_{\mu \nu} = - {1 \over 2} m_{\rm P}^2 m_h^2 h_{\mu \nu}
- {f_\phi \over 2} (\eta_{\mu \nu} \Box - \partial_\mu \partial_\nu ) \phi
+ \cdots,
\end{eqnarray}
\begin{eqnarray}
 x = c_\phi^{2} \phi + i c_\pi^2 \pi + \cdots,
\end{eqnarray}
\begin{eqnarray}
 j^\mu = m_v f_v v^\mu + f_\pi \partial^\mu \pi
+ \cdots,
\end{eqnarray}
where $\cdots$ are terms which are not linear in fields. 
The normalization of the fields are such that the propagators are given
in Appendix~\ref{app:propagators}.
We have implicitly assumed the CP invariance, {\it i.e.,} the absence of
the mixing between $\phi$ and $\pi$, for simplicity. 
By using the above parametrizations and the propagators in
Appendix~\ref{app:propagators}, one can explicitly calculate 
the correlators Eqs.~(\ref{currentCorrelatorC1})-(\ref{currentCorrelatorC8})
as a sum over the contributions from hadrons.

Following the same procedure in section~\ref{sec:GGM}, one can
derive the sum rules from $C_{1}-C_{8}$ using the effective theory. 
For example, we obtain
\begin{eqnarray}
 |f^\prime|^2 
 + |f_{\chi}|^2 
+ { 2 \over 3} |f_{\psi}|^2
=
f_\phi^2 + {8 \over 3} m_{\rm P}^2
\label{eq:sum1}
\end{eqnarray}
from the correlator $C_1$.
This rule applies to the  models with $d_{0}=3$ and $d_{0}=4$.
To derive this rule, we use two approximations;
one is the tree level approximation in the effective theory 
and the other is the perturbative calculation of the OPE for the correlator.
The effective theory should have a UV cut-off, $\Lambda_{\rm
eff}$, below which the picture of the hadron exchange (tree-level
approximation) is justified.
On the other hand, the OPE is a good expansion at a sufficiently short
distance, $(-s) > \Lambda_{\rm OPE}$, where $\Lambda_{\rm OPE}$ is a
typical scale where the UV description breaks down. 
Therefore, the above sum rule gives a good approximation 
if $\Lambda_{\rm eff} \gg \Lambda_{\rm OPE}$ and if one takes $s_0$ 
in Fig.~\ref{fig:contour} within the window, $\Lambda_{\rm OPE} < s_0 < \Lambda_{\rm eff}$.
In the case of QCD, this condition, $\Lambda_{\rm eff} > \Lambda_{\rm OPE}$,
seems to be marginally satisfied, therefore the 
Weinberg's sum rules are satisfied in the real world to a good 
accuracy.
The hadron summation in the sum rules should be taken 
while masses exceed $\Lambda_{\rm OPE}$ \cite{Shifman:1978bx,Shifman:2000jv}.

Repeating the same discussion for the rest of the correlators, $C_{2}-C_{8}$, 
we obtain sum rules:
\begin{itemize}
 \item Boson sum rule ($d_{0}=3$ and $4$)
\begin{eqnarray}
f_\phi^2 + {8 \over 3} m_{\rm P}^2 
=
f_\pi^{2} + f_v^2 ,
\label{eq:sum2}
\end{eqnarray}
 \item Scalar sum rule ($d_{2}=4$)
\begin{eqnarray}
  f_\phi c_\phi^{2} = 0, \ \ \ f_\pi c_\pi^{2} = 0,
\label{eq:sum3}
\end{eqnarray}
 \item Fermion sum rule ($d_{2}=4$)
\begin{eqnarray}
  f^2 f^\prime = m_\psi f_{\psi}^2
  = - {3 \over 4} m_\chi f_{\chi}^2.
\label{eq:sum4}
\end{eqnarray}
\end{itemize}
The correlator $C_{4}$ does not lead any sum rule for $d_{4} \le 4$.
For $d_2 > 4$  and $d_4 > 4$, there can be more sum rules. However, we
do not try to derive those in this paper since we are not aware of
such models.

\subsection{Improvement of currents and sum rules}

The entries in the sum rules, such as $f^\prime$, $f_\chi$, $f_\phi$,
and $f_\pi$, depend on the definition of the currents in the UV
theory. In deriving the sum rules, we have defined the currents as
components of the supercurrent multiplet, $J^\mu$. Moreover, we have
implicitly assumed that the current does not contain parameters with
negative mass dimensions, otherwise the dimension of ${\cal O}$ can be
arbitrarily small.

If such a supercurrent is uniquely defined, there is no
ambiguity for $f$'s. If it is not uniquely defined, the sum rules should
hold for any choice of the supercurrents.
The supercurrent $J^\mu$ has in general a freedom of the improvement,
\begin{eqnarray}
 J_\mu \to J_\mu - \partial_\mu (\Omega + \bar \Omega),
\label{eq:improve}
\end{eqnarray}
where $\Omega$ is a chiral superfield. Therefore, the improvement is
possible when there is a gauge-invariant chiral superfield with a mass
dimension less than or equal to two in the UV theory.

For example, if there is a chiral operator $M$ with dimension two and
$R$-charge zero, such as a meson operator, $M$ can be the operator
$\Omega$.  In the same way as the currents, we parametrize the
one-particle parts of the operator $M$ by low energy variables as
\begin{eqnarray}
 m &=& -{i \over \sqrt 2} 
\left( 
{F_\phi \over \sqrt 2} \phi 
- { i F_\pi \over \sqrt 2} \pi
\right) + \cdots,\\
 \psi_{M \alpha} &=& - {i \over \sqrt 2}
\left(
{ F^\prime } \lambda_\alpha 
+ {F_\chi} \chi_\alpha
\right) + \cdots,\\
 F_M &=& - i (C_\phi^{*2} \phi - i C_\pi^{*2} \pi) + \cdots,
\end{eqnarray}
where
\begin{eqnarray}
 M(y, \theta) = m(y) + \sqrt 2 \theta {\psi_M}(y) 
+ \theta \theta F_M(y).
\end{eqnarray}
With these parametrizations, the improvement in Eq.~(\ref{eq:improve}) with $\Omega = c
M$, with $c$ a real dimensionless parameter, shifts the decay constants
as
\begin{eqnarray}
 f^\prime &\to& f^\prime + c F^\prime,\\
 f_{\chi} &\to& f_{\chi} + c F_\chi,\\
 f_\phi &\to& f_\phi + c F_\phi,\\
 f_\pi &\to& f_\pi + c F_\pi,\\
 c_\phi^{2} &\to& c_\phi^{2} + c C_\phi^2,\\
 c_\pi^{2} &\to& c_\pi^{2} + c C_\pi^2.
\end{eqnarray}
The constants $f$, $f_v$, $f_{\psi}$, and $m_{\rm P}$ are unchanged by the
improvement.

When $d_0 \ge 3$, sum rules in
Eqs.~(\ref{eq:sum1}) and (\ref{eq:sum2}) should hold for any choice of
$c$. Therefore, we obtain the following relations:
\begin{eqnarray}
 |F^\prime|^2 + |F_\chi|^2 &=& F_\phi^2,
\label{eq:wave1}\\
{\rm Re}[f^{\prime *} F^\prime]
+ {\rm Re}[f_{\chi}^* F_\chi]  &=&  f_\phi F_\phi,
\label{eq:wave2}\\
 F_\phi^2 &=& F_\pi^2,
\label{eq:wave3}\\
 f_\phi F_\phi &=& f_\pi F_\pi,
\label{eq:wave4}
\end{eqnarray}
in addition to Eqs.~(\ref{eq:sum1}) and (\ref{eq:sum2}). As a trivial
example,
the effective theory described by
a single chiral superfield,
\begin{equation}
 M \propto \phi + i\pi + \sqrt{2}\theta(\lambda {\rm ~or~} \chi) + \theta\theta F,
\end{equation}
satisfies the sum rules in Eqs.~\eqref{eq:wave1}--\eqref{eq:wave4}.

\section{UV models and sum rules} 
\label{sec:UVModelsAndTheDimension}
In this section, we consider the explicit models of dynamical SUSY breaking
and discuss which sum rules in Eqs.~\eqref{eq:sum1}--\eqref{eq:sum4}  apply to them.
Here, we classify those models by whether $R$-symmetry is spontaneously
broken, and by dimensions of the SUSY breaking operators.

\subsection{Models with unbroken $\boldsymbol{R}$-symmetry}

We first discuss the models without spontaneous $R$-symmetry
breaking. In this case, the correlators with non-vanishing $R$-charges
identically vanish, and thus only Eqs.~(\ref{eq:sum1}) and
(\ref{eq:sum2}) can apply. Since $R$-symmetry is not broken, $f^\prime =
0$ in this case.
In most models, $d_0 = 4$ (except for the
model with non-vanishing $D$-term for a $U(1)$ factor), and therefore
both sum rules apply.

A famous example is the O'Raifeartaigh
model~\cite{O'Raifeartaigh:1975pr}.\footnote{There are also the
O'Raifeartaigh models with broken $R$-symmetry~\cite{Shih:2007av}.}
However, in this case, the sum rules do not give new information since
one can explicitly derive the low energy models.
Examples of dynamical SUSY breaking models are the IYIT
model~\cite{Izawa:1996pk,Intriligator:1996pu} and the ISS
model~\cite{Intriligator:2006dd} where the ISS model has unbroken
discrete $R$-symmetry.
Both of the examples have calculable IR descriptions which reduces to
the O'Raifeartaigh models.

\subsection{Models with spontaneous $\boldsymbol{R}$-symmetry breaking ($d_2 \leq 3$)}

When $R$-symmetry and SUSY are both broken by an operator with $R=2$ and
dimension less than four, those models predict the sum rules in
(\ref{eq:sum1}) and (\ref{eq:sum2}).

Examples are incalculable models such as chiral gauge theories in
Ref.~\cite{Affleck:1983vc,Affleck:1984mf}. There are also possibilities
that the incalculable K{\" a}hler potential can produce a non-trivial
$R$-symmetry breaking vacuum in the vector-like theories such as
in~\cite{Chacko:1998si, Hotta:1996ag, Ibe:2007ab}, although there are
known effective descriptions in these cases.

In models of
Ref.~\cite{Affleck:1983vc,Affleck:1984mf}, it is suggested that
the gaugino condensation, which has dimension three, breaks
both SUSY and the $R$-symmetry through the Konishi
anomaly~\cite{Konishi:1983hf}.
In the vector-like models in Ref.~\cite{Chacko:1998si,
Hotta:1996ag,Ibe:2007ab}, a dimension-three operator, $\delta_{\bar Q
\dot \alpha} ( \bar \psi_S^{\dot\alpha} S )$, is the one which breaks
both SUSY and $R$-symmetry, where $\psi_S$ and $S$ are the fermionic and the
bosonic components of a gauge singlet chiral superfield.

\subsection{Models with spontaneous $\boldsymbol{R}$-symmetry breaking ($d_2 \geq 4$)}

Possibly some gauge theory without a matter field can be of this type,
although there is no known example.
In this case, all the sum rules in Eqs.~(\ref{eq:sum1})--(\ref{eq:sum4}) can
be derived.

Since $R$-symmetry is spontaneously broken, one can say $f_\pi \neq
0$. This implies that the left-hand side of Eq.~\eqref{eq:sum2} is
non-vanishing and therefore the left-hand side of Eq.~\eqref{eq:sum1} is
also non-vanishing. Together with Eq.~\eqref{eq:sum4}, $m_\psi f_\psi^2$
is non-vanishing (unless there is a cancellation among same-spin
fermions). Therefore, this type of model generally involves massive
spin-3/2 field.

If one finds that the sum rules in Eqs.~(\ref{eq:sum3}) and
(\ref{eq:sum4}) apply in some hadronic models of SUSY breaking such as
the dual gravity constructions~\cite{Kachru:2002gs, Argurio:2006ny,
Argurio:2007qk}, it may be suggesting that the microscopic description
is in this category.

\section{Discussions}

We have derived sum rules for hadrons in dynamical SUSY breaking models.
The sum rules involve massive fields with spin 3/2 and 2. It is
interesting to note here that there is an analogy of this situation in
QCD.

The Nambu-Goldstone bosons (pions) associated with the chiral symmetry
breaking are described by a non-linear sigma model (chiral Lagrangian)
which has a UV cut-off scale.
The cut-off scale can be pushed higher by including massive hadrons.
The simplest possibility is to promote the non-linear sigma model to a
linear-sigma one by introducing a scalar field (which is usually called
the sigma meson). 
However, the actual hadronic world did not choose that realization,
instead, a vector meson (the rho meson) appeared as the next lightest
state. 
In view of such situation, the Hidden Local Symmetry (HLS)
model~\cite{Bando:1984ej} is proposed, in which the rho meson is
introduced as a massive vector boson of a hidden local SU(2) symmetry.

In SUSY breaking case, the low-energy effective Lagrangian is formulated
by Volkov and Akulov in Ref.~\cite{Volkov:1972jx}, where the
Nambu-Goldstone fermion, the Goldstino, is introduced as non-linearly
transforming field under SUSY.
The simplest possibility for the next lightest mode is the superpartner
of the Goldstino, formulating the low-energy effective model with a
chiral supermultiplet. This is analogous to the linear sigma model
realization of the chiral symmetry case. 
As in QCD, it is worth considering an alternative realization, namely
the SUSY breaking model equivalent of the HLS realization.  Such a
realization is achieved by introducing the massive spin-2 field, as
discussed in Ref.~\cite{Graesser:2009bu}.

Another realization of the massive higher spin states in SUSY gauge
theories is related to the gauge/gravity correspondence.  For example,
in the Holographic QCD model \cite{Son:2003et, Sakai:2004cn,
Erlich:2005qh, Da Rold:2005zs}, the HLS naturally emerges and the rho
meson appears as a ``Kaluza-Klein (KK)'' excitation mode of the
five-dimensional gauge field in the holographic dual.  In the context of
the gauge/gravity duality, the possibility of the dynamical SUSY
breaking has been discussed~\cite{Kachru:2002gs, Argurio:2006ny,
Argurio:2007qk}.  If the gravity dual of the dynamical SUSY breaking
model is successfully constructed, the Goldstino should be identified
with a normalizable zero mode of the KK modes of the bulk gravitino
\cite{Argurio:2006my}.  Furthermore, massive spin-3/2 and massive spin-2
modes also appear from gravitino and graviton in the dual supergravity.
In this sense, our effective theory with the hidden local SUSY can be
related to the dual supergravity.

\section*{Acknowledgements}
We thank Noriaki Kitazawa for conversation on the Weinberg sum rules.
RK also thanks Matthew Sudano for valuable discussions. RK is supported
in part by the Grant-in-Aid for Scientific Research 21840006 and
23740165 of JSPS. NM is supported by the GCOE program 
``Weaving Science Web beyond Particle-Matter Hierarchy.''

\appendix
\section{Propagators}
\label{app:propagators}

\begin{eqnarray}
 \langle \lambda_\alpha (x)
\bar \lambda_{\dot \beta} (y)
\rangle
= {1 \over f^4} \ (\sigma^\rho)_{\alpha \dot \beta}
\int {d^4 k \over i (2 \pi)^4}
{k_\rho \over - k^2 - i \epsilon }
e^{-i k \cdot (x-y)}.
\end{eqnarray}
\begin{eqnarray}
 \langle \chi_\alpha (x)
\bar \chi_{\dot \beta} (y)
\rangle
= (\sigma^\rho)_{\alpha \dot \beta}
\int {d^4 k \over i (2 \pi)^4}
{k_\rho \over m_\xi^2 - k^2 - i \epsilon }
e^{-i k \cdot (x-y)}.
\end{eqnarray}
\begin{eqnarray}
 \langle \chi_\alpha (x)
 \chi^\beta (y)
\rangle
= \delta_\alpha^\beta
\int {d^4 k \over i (2 \pi)^4}
{m_\chi \over m_\chi^2 - k^2 - i \epsilon }
e^{-i k \cdot (x-y)}.
\end{eqnarray}
\begin{eqnarray}
 \langle 
\psi_{\mu \alpha} (x)
\bar \psi_{\nu \dot \beta} (y)
\rangle
= \left(
P_L \langle 
\Psi_\mu (x) \bar \Psi_\nu (y)
\rangle P_R
\right)_{\alpha \dot \beta}.
\end{eqnarray}
\begin{eqnarray}
 \langle 
\psi_{\mu \alpha} (x)
\psi_{\nu}^\beta (y)
\rangle
= \left(
P_L \langle 
\Psi_\mu (x) \bar \Psi_\nu (y)
\rangle P_L
\right)_{\alpha}^{\ \beta}.
\end{eqnarray}
\begin{eqnarray}
 \langle \Psi_\mu (x) \bar \Psi_\nu (y) \rangle
= \int {d^4 k \over i (2 \pi)^4 }
{P_{\mu \nu} (k) \over m_\psi^2 - k^2 - i \epsilon}
e^{-i k \cdot (x-y)}.
\end{eqnarray}
\begin{eqnarray}
 P_{\mu \nu}
= 
- \left(
\eta_{\mu \nu} - {k_\mu k_\nu \over m_\psi^2} 
\right)
\left(
\slashed k + m_\psi
\right)
-{1 \over 3}
\left(
\gamma_\mu + {k_\mu \over m_\psi}
\right)
(\slashed k - m_\psi )
\left(
\gamma_\nu + {k_\nu \over m_\psi}
\right).
\end{eqnarray}
\begin{eqnarray}
 \langle h^{\mu \nu} (x) h^{\rho \sigma} (y) \rangle
= {2 \over m_{\rm P}^2}
\int {d^4 k \over i (2 \pi)^4 }
{ B^{\mu \nu ; \rho \sigma} \over m_h^2 - k^2 - i \epsilon}
e^{-i k \cdot (x-y)}.
\end{eqnarray}
\begin{eqnarray}
 B_{\mu \nu ; \rho \sigma} 
&=& \left(
\eta_{\mu \rho} - {k_\mu k_\rho \over m_h^2}
\right)
\left(
\eta_{\nu \sigma} - {k_\nu k_\sigma \over m_h^2}
\right)
+
\left(
\eta_{\mu \sigma} - {k_\mu k_\sigma \over m_h^2}
\right)
\left(
\eta_{\nu \rho} - {k_\nu k_\rho \over m_h^2}
\right)
\nonumber \\
&&
- {2 \over 3}
\left(
\eta_{\mu \nu} - {k_\mu k_\nu \over m_h^2}
\right)
\left(
\eta_{\rho \sigma} - {k_\rho k_\sigma \over m_h^2}
\right).
\end{eqnarray}
\begin{eqnarray}
 \langle \phi (x) \phi (y) \rangle
= \int {d^4 k \over i (2 \pi)^4 }
{1 \over m_\phi^2 - k^2 - i \epsilon}
e^{-i k \cdot (x-y)}.
\end{eqnarray}
\begin{eqnarray}
 \langle \pi (x) \pi (y) \rangle
= \int {d^4 k \over i (2 \pi)^4 }
{1 \over m_\pi^2 - k^2 - i \epsilon}
e^{-i k \cdot (x-y)}.
\end{eqnarray}
\begin{eqnarray}
 \langle v^\mu (x) v^\nu (y) \rangle
= \int {d^4 k \over i (2 \pi)^4 }
{\eta^{\mu \nu} - {k^\mu k^\nu / m_v^2}
\over k^2 - m_v^2 + i \epsilon} 
e^{-i k \cdot (x-y)}.
\end{eqnarray}

\end{document}